\documentclass[prb,twocolumn,showpacs]{revtex4}
\usepackage{graphicx,epsf}

\newcommand{\bq}{{\bf q}}
\newcommand{\bk}{{\bf k}}

\newcommand{\bR}{{\bf R}}

\newcommand{\ptilde}{\tilde{p}}
\newcommand{\stilde}{\tilde{s}}
\newcommand{\ctilde}{\tilde{c}}
\newcommand{\ltilde}{\tilde{l}}
\newcommand{\vtilde}{\tilde{v}}
\newcommand{\rhobar}{\bar{\rho}}
\newcommand{\chibar}{\bar{\chi}}
\newcommand{\rhobarbar}{\bar{\bar{\rho}}}
\newcommand{\chibarbar}{\bar{\bar{\chi}}}
\newcommand{\Hhat}{\hat{H}}
\newcommand{\Htilde}{\tilde{H}}

\begin{document}

\def\tende#1{\,\vtop{\ialign{##\crcr\rightarrowfill\crcr
\noalign{\kern-1pt\nointerlineskip}
\hskip3.pt${\scriptstyle #1}$\hskip3.pt\crcr}}\,}

\title{Second generation of composite fermions in the Hamiltonian theory}
\author{M.\ O.\ Goerbig$^{1,2}$, P.\ Lederer$^2$, and  
C.\ Morais\ Smith$^1$}

\affiliation{$^1$D\'epartement de Physique, Universit\'e de Fribourg, P\'erolles,  CH-1700 Fribourg, Switzerland.\\
$^2$Laboratoire de Physique des Solides, Bat.\,510, UPS (associ\'e au CNRS), F-91405 Orsay cedex, France.}

\begin{abstract}
In the framework of a recently developed model of interacting composite 
fermions restricted to a single level, we calculate the activation gaps of a
second generation of spin-polarized composite fermions. These composite 
particles consist each of a composite fermion of the first generation and a 
vortex-like excitation and may be responsible for the recently observed 
fractional quantum Hall states at unusual filling factors such as 
$\nu=4/11,5/13,5/17$, and $6/17$. Because the gaps of composite fermions of 
the second generation are found to be more than one order of magnitude smaller
than those of the first generation, these states are less visible than the 
usual states observed at filling factors $\nu=p/(2ps+1)$. Their stability is 
discussed in the context of a pseudopotential expansion of the 
composite-fermion interaction potential. 
\end{abstract}
\maketitle

\section{Introduction}

The fractional quantum Hall effect (FQHE) in the lowest Landau level
(LL) at filling factors $\nu=p/(2ps+1)$, with integral $p$ and $s$,
may be understood as an integral quantum Hall effect (IQHE) in terms
of composite fermions (CFs). \cite{jain,perspectives} Each CF consists of a
bound state of one electron and a vortex-like excitation with
vorticity $2s$ and charge $c^2=2ps/(2ps+1)$, in units of the electron
charge. The CFs experience a reduced coupling $(eB)^*=eB/(2ps+1)$ to
the external magnetic field $B$ and thus populate CF-LLs with a CF
filling factor $\nu^*=n_{el}/n_{B^*}$. Here, $n_{el}$ is the
electronic density, and $n_{B^*}=1/2\pi l_B^{*2}$ is the 
density of states of each CF-LL in terms of the CF magnetic length 
$l_B^{*}=\sqrt{\hbar/(eB)^*}$. The CF and electronic filling factors are
related by $\nu=\nu^*/(2s\nu^*+1)$.
The FQHE arises when $\nu^*=p$: because of the CF-LL separation, which
is on the order of the characteristic Coulomb energy $e^2/\epsilon
l_B$, a finite energy is required to promote a particle to an upper
CF-LL if the filling is changed. These particles become localized by
residual impurities and therefore do not contribute to the electrical
transport, giving rise to the observed plateaus in the Hall resistance.

Experiments by Pan {\sl et al.} have revealed a new class of FQHE
states at filling factors such as $\nu=4/11,6/17,...$, of which at
least the 4/11 state is found to be spin-polarized. \cite{pan} The existence of
such states had been conjectured by Mani and v.\ Klitzing based on the
fractal structure and self-similarity of the Hall resistance
curve. \cite{mani} It appears natural to interpret these states, found
at CFfilling factors $\nu^*=1+\ptilde/(2\stilde\ptilde+1)$, in terms of
an IQHE of a second generation of CFs (C$^2$Fs); whereas the CFs 
in the lowest CF-LL ($p=0$) remain inert, those in the partially filled $p=1$ 
level bind to some vortex-like excitation carrying $2\stilde$ additional flux 
quanta. This picture is reminiscent of the hierarchy scheme proposed by 
Haldane and Halperin, according to which Laughlin quasiparticles 
\cite{laughlin} might in principle form an incompressible liquid state if
the interaction potential between these quasiparticles is sufficiently 
short-range. \cite{haldane,halperin} Even though the new states may 
be classified within this scheme, theoretical studies seem to indicate that
fully spin-polarized
C$^2$Fs and even higher generations of CFs would not be stable. 
\cite{beran,wojs,mandaljain,chang} Exact-diagonalization studies
for the $4/11$ state show a gapped ground state only 
for $12$ electrons, which is at present the maximum electron number for
such numerical investigations. \cite{morfPC} On the other hand, a numerical
diagonalization with up to $24$ 
particles can be performed in the CF basis with a truncated Hilbert space. 
\cite{mandaljain,chang} Within the latter formalism, a 
ground state with features of a FQHE state emerges for $N=12$ and $20$
particles, but not 
for $N=8,16$, and $24$. \cite{mandaljain} The extrapolation to the 
thermodynamic limit is therefore ambiguous. Lopez and Fradkin
investigated the topological stability of a FQHE state at $\nu=4/11$
in a Chern-Simons field-theory approach. \cite{lopezfradkin} In
contrast to prior stability investigations by Haldane, \cite{haldane2}
a spin-polarized state is found to be {\sl topologically stable}, if 
interpreted in terms of C$^2$Fs.

Recently, we have derived a model of interacting CFs \cite{goerbigUP} in the 
framework of the Hamiltonian theory proposed by Murthy and Shankar. \cite{MS} 
The involved transformations provide a mathematical basis for the 
self-similarity of the FQHE. \cite{goerbigUP} Here, we construct
a C$^2$F representation of this model, in analogy with the CF representation 
in the Hamiltonian theory, by the restriction of the
dynamics to the first excited CF-LL $p=1$. We suppose the stability
of fully spin-polarized C$^2$F states and calculate their activation gaps 
as a function of the finite width of the two-dimensional electron system.
In contrast to prior numerical analyses in the wave-function approach,
the calculations are performed directly in the thermodynamic limit. 
Inter-CF-LL excitations are taken into account and give rise to a screened 
effective interaction. The dielectric function, which is calculated in the 
random-phase approximation (RPA), is similar to the electronic case discussed
by Aleiner and Glazman. \cite{AG} 

The structure of the paper is the following:
in Sec. II we present the model of interacting CFs and evaluate their
effective 
interaction potential. The dielectric function due to inter-CF-LL excitations
is derived in Sec. III. In Sec. IV we construct the C$^2$F basis and
calculate the activation gaps of the C$^2$F states as a function
of the finite width of the electron system. The resulting activation gap 
for $\nu=4/11$ is compared to the experimental estimate. The stability of 
C$^2$F states is discussed in terms of a pseudopotential expansion of the 
CF interaction potential in Sec. V.  A brief summary may be found in Sec. VI.

\section{Model}

We adopt a model of spin-polarized particles, the dynamics of which are 
restricted to the lowest electronic 
LL. The low-energy degrees of freedom are given by the Hamiltonian
\begin{equation}
\label{equ001}
\Hhat=\frac{1}{2}\sum_{\bq}v_0(q)\rhobar(-\bq)\rhobar(\bq),
\end{equation}
where $v_0(q)=(2\pi e^2/\epsilon q)\exp(-q^2l_B^2/2)$, with 
$l_B=\sqrt{\hbar/eB}$. Instead of representing the projected density operators
$\rhobar(\bq)$ in the usual electron basis, an alternative representation in
terms of CFs is chosen in the Hamiltonian theory of the FQHE: a ``preferred''
combination of the projected density and the density $\chibar(\bq)$ of the 
vortex-like object is introduced, 
$\rhobar_{CF}(\bq)=\rhobar(\bq)-c^2\chibar(\bq)$.\cite{MS} This preferred
combination plays the role of the CF density operator and may be expressed as
\begin{eqnarray}
\nonumber
\rhobar_{CF}(\bq)&=&\sum_{{n,n'};{m,m'}}\langle m|e^{-i\bq\cdot\bR_{CF}}|m'\rangle \\
\nonumber
&&\times \langle n|\rhobar^{p}(\bq)|n'\rangle c_{n,m}^{\dagger}c_{n',m'},
\end{eqnarray}

\noindent
where $c_{n,m}^{\dagger}$ creates a CF in the $n$-th CF-LL with the
guiding-center quantum number $m$. The matrix elements are (for $m\geq m'$)
\begin{eqnarray}
\label{equ004}
\nonumber
\langle
m|e^{-i\bq\cdot\bR_{CF}}|m'\rangle&=&\sqrt{\frac{m'!}{m!}}\left(\frac{-i(q_x+iq_y)l_B^*}{\sqrt{2}}\right)^{m-m'}\\
&&L_{m'}^{m-m'}\left(\frac{q^2l_B^{*2}}{2}\right)e^{-q^2l_B^{*2}/4},
\end{eqnarray}
and (for $n\geq n'$)
{\small
\begin{eqnarray}
\label{equ007}
\nonumber
\langle n|\rhobar^p(\bq)|n'\rangle&=&\sqrt{\frac{n'!}{n!}}\left(\frac{-i(q_x-iq_y)l_B^{*}c}{\sqrt{2}}\right)^{n-n'}e^{-q^2l_B^{*2}c^2/4}\\
&&\times\left[L_{n'}^{n-n'}\left(\frac{q^2l_B^{*2}c^2}{2}\right)\right.\\
\nonumber
&&\left.-c^{2(1-n+n')}e^{-q^2l_B^2/2c^2}L_{n'}^{n-n'}\left(\frac{q^2l_B^{*2}}{2c^2}\right)\right],
\end{eqnarray}
}

\noindent
with the associated Laguerre polynomials $L_n^m(x)$ and the CF magnetic length
$l_B^*=l_B/\sqrt{1-c^2}$. 

In order to describe the low energy degrees of freedom in the experimentally 
relevant range of CF fillings $1<\nu^*<2$, we restrict $\rhobar_{CF}(\bq)$
to the CF-LL $p=1$ in the same manner as for electrons. The restricted 
CF-density operator is
$\langle \rhobar_{CF}(\bq)\rangle_{p=1}=\langle1|\rhobar^p(\bq)|1\rangle \rhobarbar (\bq)$, with 
$$
\rhobarbar (\bq)=\sum_{m,m'}\langle m|e^{-i\bq\cdot\bR_{CF}}|m'\rangle c_{n=1,m}^{\dagger}c_{n=1,m'}.
$$
The model Hamiltonian of restricted CFs therefore becomes\cite{goerbigUP}
\begin{equation}
\label{equ011}
\Htilde(s)=\frac{1}{2}\sum_{\bq}\vtilde(q)\rhobarbar(-\bq)\rhobarbar(\bq),
\end{equation}
where the CF form factor 
\begin{eqnarray}
\nonumber
F_{CF}(q)&\equiv&\langle 1|\rhobar^p(\bq)|1\rangle =e^{-q^2l_B^{*2}c^2/4}\left[L_1\left(\frac{q^2l_B^{*2}c^2}{2}\right)\right.\\
\nonumber
&&\left.\qquad\qquad -c^2e^{-q^2l_B^2/2c^2}L_1\left(\frac{q^2l_B^{*2}}{2c^2}\right)\right]
\end{eqnarray}
has been absorbed into an effective CF interaction potential
\begin{equation}
\label{equ012}
\vtilde(q)=v_0(q)\frac{[F_{CF}(q)]^2}{\epsilon(q)}.
\end{equation}
Note that the derivation of this CF model is not limited to $p=1$, but may
also be applied for higher CF-LLs. In this case the CF form factor depends on 
$p$, $F_{CF}^p(q)=\langle p|\rhobar^p(\bq)|p\rangle$. The projected CF-density
operators satisfy the same commutation relations 
$$[\rhobarbar(\bq),\rhobarbar(\bk)]=2i\sin\left(\frac{(\bq\times\bk)_zl_B^{*2}}{2}\right)\rhobarbar(\bq+\bk)$$
as the projected density operators in the model of electrons restricted to 
the lowest LL if one replaces the electronic by the CF magnetic 
length.\cite{MS,GMP} This and the fact that the Hamiltonian of interacting CFs
(\ref{equ011}) has the same structure as the original electronic
one\cite{goerbigUP} allows us treat the model with the same
theoretical tools which 
have been used for the understanding of the FQHE at $\nu=p/(2ps+1)$. One
would simply need to replace the electronic by the CF interaction potential,
$v_0(q)\rightarrow \vtilde(q)$, and take into account the renormalization of
the magnetic length, $l_B\rightarrow l_B^*$. 

\section{Dielectric function}

In contrast to the
electronic case, inter-CF-LL excitations have to be considered
explicitly in a dielectric function $\epsilon(q)$, which modifies the
CF interaction potential, because the residual CF interactions are 
intrinsically on the same order of magnitude as the CF-LL separation - 
both energy scales are given by the Coulomb interaction
$e^2/\epsilon l_B$.  

For the calculation of the dielectric function, we investigate inter-CF-LL
excitations in the original Hamiltonian (\ref{equ001}) in the CF basis. Its
solution in the Hartree-Fock approximation yields the ``free'' CF Hamiltonian
\begin{equation}
\label{equ014}
\Hhat_{CF}^0=\sum_{n,m}\varepsilon_n c_{n,m}^{\dagger}c_{n,m}.
\end{equation}
It can be shown that the CF-LLs are approximately linear in $n$, 
$\epsilon_n\simeq n\omega_C^*$, where $\omega_C^*$ is the 
energy of a quasiparticle excitation in units of $\hbar\equiv 1$,
and we have omitted an unimportant constant. The 
corrections beyond the Hartree-Fock approximation are given by the 
``interaction'' Hamiltonian
\begin{eqnarray}
\label{equ015}
\nonumber
\Hhat_{CF}^{int}&=&\frac{1}{2}\sum_{\nu_1,...,\nu_4}v_{\nu_1,...,\nu_4}\left[c_{\nu_1}^{\dagger}c_{\nu_2}^{\dagger}c_{\nu_4}c_{\nu_3}\right.\\
\nonumber
&&-\langle c_{\nu_1}^{\dagger}c_{\nu_3}\rangle
c_{\nu_2}^{\dagger}c_{\nu_4}-\langle
c_{\nu_2}^{\dagger}c_{\nu_4}\rangle c_{\nu_1}^{\dagger}c_{\nu_3}\\
&&\left. +\langle c_{\nu_1}^{\dagger}c_{\nu_4}\rangle
c_{\nu_2}^{\dagger}c_{\nu_3} +\langle c_{\nu_2}^{\dagger}c_{\nu_3}\rangle
c_{\nu_1}^{\dagger}c_{\nu_4}\right],
\end{eqnarray}
with $\nu_i=(n_i,m_i)$ and the interaction vertex 
\begin{eqnarray}
\nonumber
v_{\nu_1,...,\nu_4}&=&\sum_{\bq}v_0(q)\langle
n_1|\rhobar^p(-\bq)|n_3\rangle\langle n_2|\rhobar^p(\bq)|n_4\rangle \\
\nonumber
&& \langle m_1|e^{i\bq\cdot\bR_{CF}}|m_3\rangle \langle
m_2|e^{-i\bq\cdot\bR_{CF}}|m_4\rangle .
\end{eqnarray}
The special form of the interaction Hamiltonian (\ref{equ015}) leads
to the omission of all diagrams containing the equal-time contractions
$\langle \mathcal{T}c_{n,m}(\tau)c_{n,m}^{\dagger}(\tau)\rangle$ in a
diagrammatic perturbation expansion, where $\mathcal{T}$ denotes time-ordering.
The averages are taken over the 
Hartree-Fock ground state characterized by $\langle c_{n,m}^{\dagger}c_{n',m'}\rangle=\delta_{n,n'}\delta_{m,m'}\Theta(p-1-n)$, where $\Theta(x)$ is
the step function with $\Theta(x)=0$ for $x<0$ and $\Theta(x)=1$ for $x\geq0$.
The dielectric function may be
calculated in the RPA,
$\epsilon_{RPA}(\bq,\omega)=1-v_0(q)D(\bq,\omega),$
where
\begin{eqnarray}
\nonumber
D(\bq,\omega)&=&-i\int dt e^{i\omega t}\langle
\mathcal{T}\rhobar_{CF}(\bq,t)\rhobar_{CF}(-\bq,t=0)\rangle\\
\nonumber
&=&2\sum_{n>0}\mathcal{F}_n^p(q)\frac{n\omega_C^*}{\omega^2-n^2\omega_C^{*2}}
\end{eqnarray}
is the density-density Green's function with
\begin{equation}
\label{equ019}
\mathcal{F}_n^p(q)=\sum_{n'=p-n}^{p-1}|\langle
n'+n|\rhobar^p(\bq)|n'\rangle|^2.
\end{equation}
For lower $p$, the static dielectric function in the limit 
$\omega\rightarrow0$ is a finite sum over Laguerre polynomials, which may be 
directly calculated with the help of the matrix elements (\ref{equ007}). 
In the case of $p=1$ one finds
\begin{eqnarray}
\nonumber
\epsilon_{RPA}^{(p=1)}(q)&=&1+\frac{v_0(q)(ql_B^*c)^2}{\omega_C^*}\left[L_1\left(\frac{q^2l_B^{*2}c^2}{2}\right)\right.\\
\nonumber
&&-\left.e^{-q^2l_B^2/2c^2}L_1\left(\frac{q^2l_B^{*2}}{2c^2}\right)\right]^2 e^{-q^2l_B^{*2}c^2/2}.
\end{eqnarray}
For larger values of $p$, one may use the asymptotic form of the amplitudes 
(\ref{equ019})
$$\mathcal{F}_n^p(q)\simeq n\left[J_n(ql_B^*c\sqrt{2p+1})-c^2J_n\left(\frac{ql_B^*}{c}\sqrt{2p+1}\right)\right]^2,$$
in terms of the Bessel functions $J_n(x)$. This yields the static dielectric 
function
\begin{eqnarray}
\nonumber
\label{equ020}
\epsilon_{RPA}^p(q)&\simeq&1+\frac{v_0(q)}{\omega_C^*}\left\{1+c^4-2c^2
  J_0\left[qR_C^*\left(c-\frac{1}{c}\right)\right]\right.\\
&&-\left.\left[J_0(qR_C^*c)-c^2J_0\left(\frac{qR_C^*}{c}\right)\right]^2\right\},
\end{eqnarray}
with the CF cyclotron radius $R_C^*=l_B^*\sqrt{2p+1}$.
This is similar to the sreening of the effective electron-electron
interaction in higher electronic LLs. \cite{AG} Note that the RPA is more 
problematic in the CF than in the electronic case because 
$(e^2/\epsilon l_B)/\omega_C^*$ is not a physically small parameter. 
Therefore, other diagrams may also play a role in the dielectric function.

\section{C$^2$F basis and activation gaps}

In order to obtain the C$^2$F representation of the model, we proceed
in the same manner as for the construction of the CF representation. \cite{MS} 
A new preferred combination is introduced,
$\rhobar_{C^2F}(\bq)=\rhobarbar(\bq)-\ctilde^2\chibarbar(\bq)$, where
$\chibarbar(\bq)$ describes a vortex-like excitation of the CF  
liquid with vorticity $2\stilde$ and charge
$\ctilde^2=2\ptilde\stilde/(2\ptilde\stilde+1)$ in units of the CF
charge $e^*=1-c^2$. This density operator becomes in the C$^2$F basis
\begin{eqnarray}
\nonumber
\rhobar_{C^2F}(\bq)&=&\sum_{n,n';m,m'}\langle
m|e^{-i\bq\cdot\bR_{C^2F}}|m'\rangle \\
\nonumber
&&\times\,\langle n|\rhobarbar^{p}(\bq)|n'\rangle
\ctilde_{n,m}^{\dagger}\ctilde_{n',m'},
\end{eqnarray} 
where $\ctilde_{n,m}^{\dagger}$ creates a C$^2$F in the C$^2$F-LL $n$
in the guiding-center state $m$. The degeneracy of each C$^2$F-LL is
now $n_{\tilde{B}}=1/2\pi\ltilde^2$, where
$\ltilde=l_B^*/\sqrt{1-\ctilde^2}=l_B/\sqrt{(1-c^2)(1-\ctilde^2)}$ is
the C$^2$F magnetic length. The matrix elements
are given by Eqs.\,(\ref{equ004}) and (\ref{equ007}) if one replaces 
$l_{B}^*\rightarrow \ltilde$ and $c\rightarrow\ctilde$. 

The activation gap for C$^2$Fs is calculated in the same manner as for
CFs; \cite{MS} it is the sum of the energies of a quasiparticle and a 
quasihole excitation
$\Delta^a(s;\stilde,\ptilde)=\Delta^{qp}(s;\stilde,\ptilde)+\Delta^{qh}(s;\stilde,\ptilde)$,
with 
$$\Delta^{qp}(s;\stilde,\ptilde)=\langle \ctilde_{\ptilde,m}\Htilde(s)\ctilde_{\ptilde,m}^{\dagger}\rangle-\langle \Htilde\rangle$$ 
and 
$$\Delta^{qh}(s;\stilde,\ptilde)=\langle \ctilde_{\ptilde-1,m}^{\dagger}\Htilde(s)\ctilde_{\ptilde-1,m}\rangle-\langle \Htilde\rangle.$$ 
With the help of Wick contractions, one averages over
the C$^2$F ground state, which consists of $\ptilde$ completely filled 
C$^2$F-LLs and may thus be characterized by $\langle
\ctilde_{n,m}^{\dagger}\ctilde_{n',m'}\rangle=\delta_{n,n'}\delta_{m,m'}\Theta(\ptilde-1-n)$.
This yields the activation gaps
\begin{eqnarray}
\label{equ13}
\nonumber
\Delta^a(s;\stilde,\ptilde)=\frac{1}{2}\sum_{\bq}\vtilde(q)\left[\langle\ptilde|\rhobarbar^p(-\bq)\rhobarbar^p(\bq)|\ptilde\rangle \right.\\
\nonumber
\left.\qquad 
-\langle\ptilde-1|\rhobarbar^p(-\bq)\rhobarbar^p(\bq)|\ptilde-1\rangle\right]\\
\nonumber
-\sum_{\bq}\vtilde(q)\sum_{n=0}^{\ptilde-1}\left[|\langle\ptilde|\rhobarbar^p(\bq)|n\rangle|^2 
- |\langle\ptilde-1|\rhobarbar^p(\bq)|n\rangle|^2\right],
\end{eqnarray}
in analogy with the case of CFs. \cite{MS}

In units of $e^2/\epsilon l_B$ and for characteristic quasiparticle
energies $\omega_C^*$ found in the litterature,\cite{jain,MS,morf} the 
activation gaps of some chosen states with $s=1$ and 
$p=1$ (with $\omega_C^*=0.1$) read

\vspace{0.2cm}
\begin{tabular}{|c||c|c|}
\hline
$\stilde=1$ & $\ptilde=1$ & $\ptilde=2$  \\ \hline
$\nu^*$ & $1+1/3$ & $1+2/5$ \\ \hline
$\nu$ & $4/11$ & $7/19$  \\ \hline \hline
$\Delta^a$ & 0.0067 & 0.0027  \\ \hline
\end{tabular}\hspace{0.5cm}
\begin{tabular}{|c||c|c|}
\hline
$\stilde=2$ & $\ptilde=1$ & $\ptilde=2$  \\ \hline
$\nu^*$ & $1+1/5$ & $1+2/9$  \\ \hline
$\nu$ & $6/17$ & $11/31$ \\ \hline \hline
$\Delta^a$ & 0.0031 & 0.0013 \\ \hline
\end{tabular}
\vspace{0.2cm}

\noindent
whereas for states with $s=2$ and $p=1$ (with $\omega_C^*=0.03$), one obtains

\vspace{0cm}
\begin{center}
\begin{tabular}{|c||c|c|}
\hline
$\stilde=1$ & $\ptilde=1$ & $\ptilde=2$ \\ \hline
$\nu^*$ & $1+1/3$ & $1+2/5$ \\ \hline
$\nu$ & $4/19$ & $7/33$ \\ \hline \hline
$\Delta^a$ & 0.0016 & 0.00070 \\ \hline
\end{tabular}\hspace{0.5cm}
\begin{tabular}{|c||c|c|}
\hline
$\stilde=2$ & $\ptilde=1$ & $\ptilde=2$ \\ \hline
$\nu^*$ & $1+1/5$ & $1+2/9$ \\ \hline
$\nu$ & $6/29$ & $11/53$ \\ \hline \hline
$\Delta^a$ & 0.00082 & 0.00035  \\ \hline
\end{tabular}
\end{center}
\vspace{0cm}

\noindent
Note that the C$^2$F states at 
$\nu^*=1+\ptilde/(2\ptilde\stilde+1)$ and 
$\nu^*=2-\ptilde/(2\ptilde\stilde+1)$ are related by the particle-hole
symmetry in the same manner as the CF states at $\nu=p/(2ps+1)$ and
$\nu=1-p/(2ps+1)$. The $4/11$ state is therefore equivalent to
the states at $\nu=5/13,7/11$, and $8/13$. 

\begin{figure}
\epsfysize+5.2cm
\epsffile{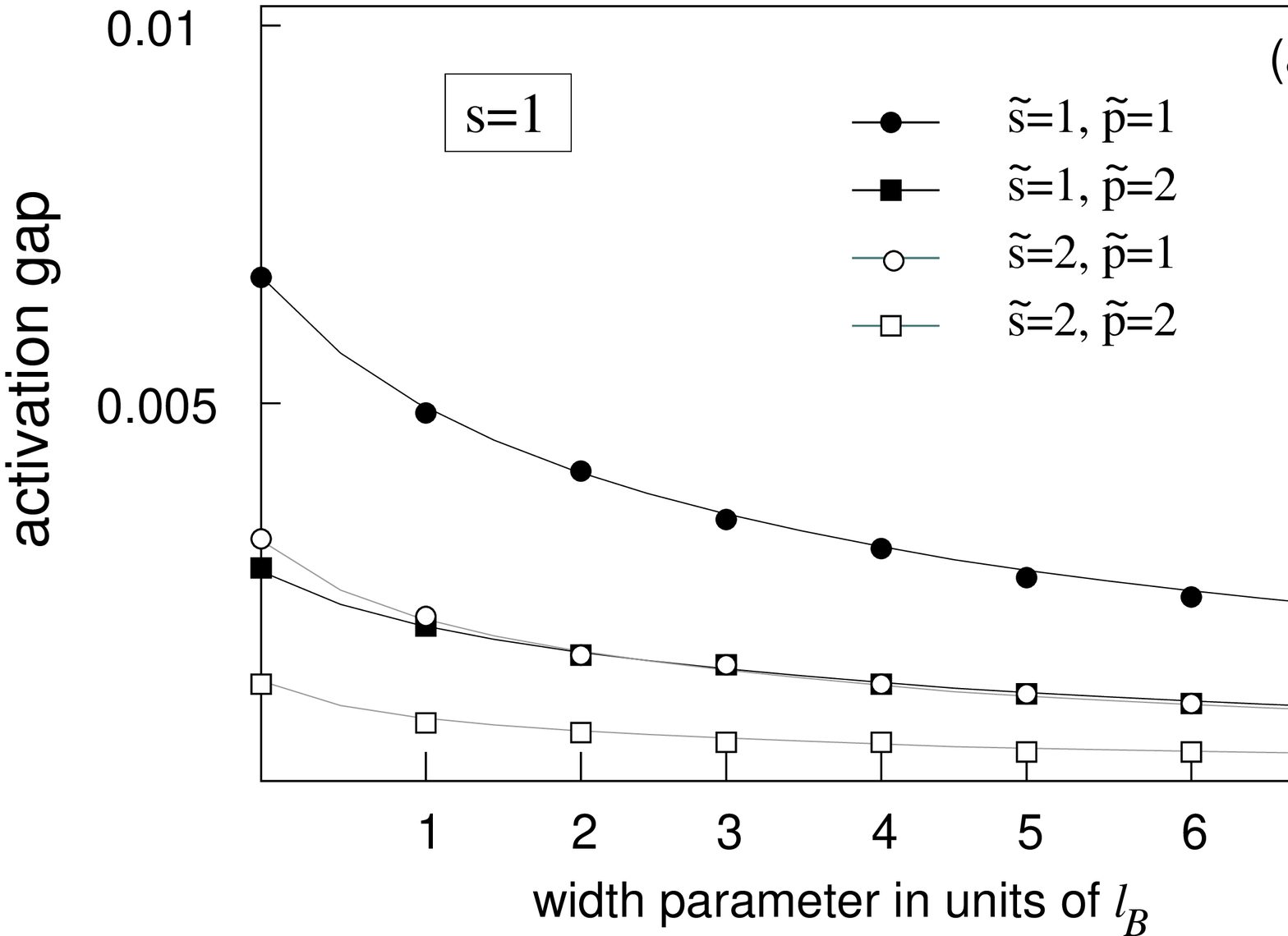}
\epsfysize+5.6cm
\epsffile{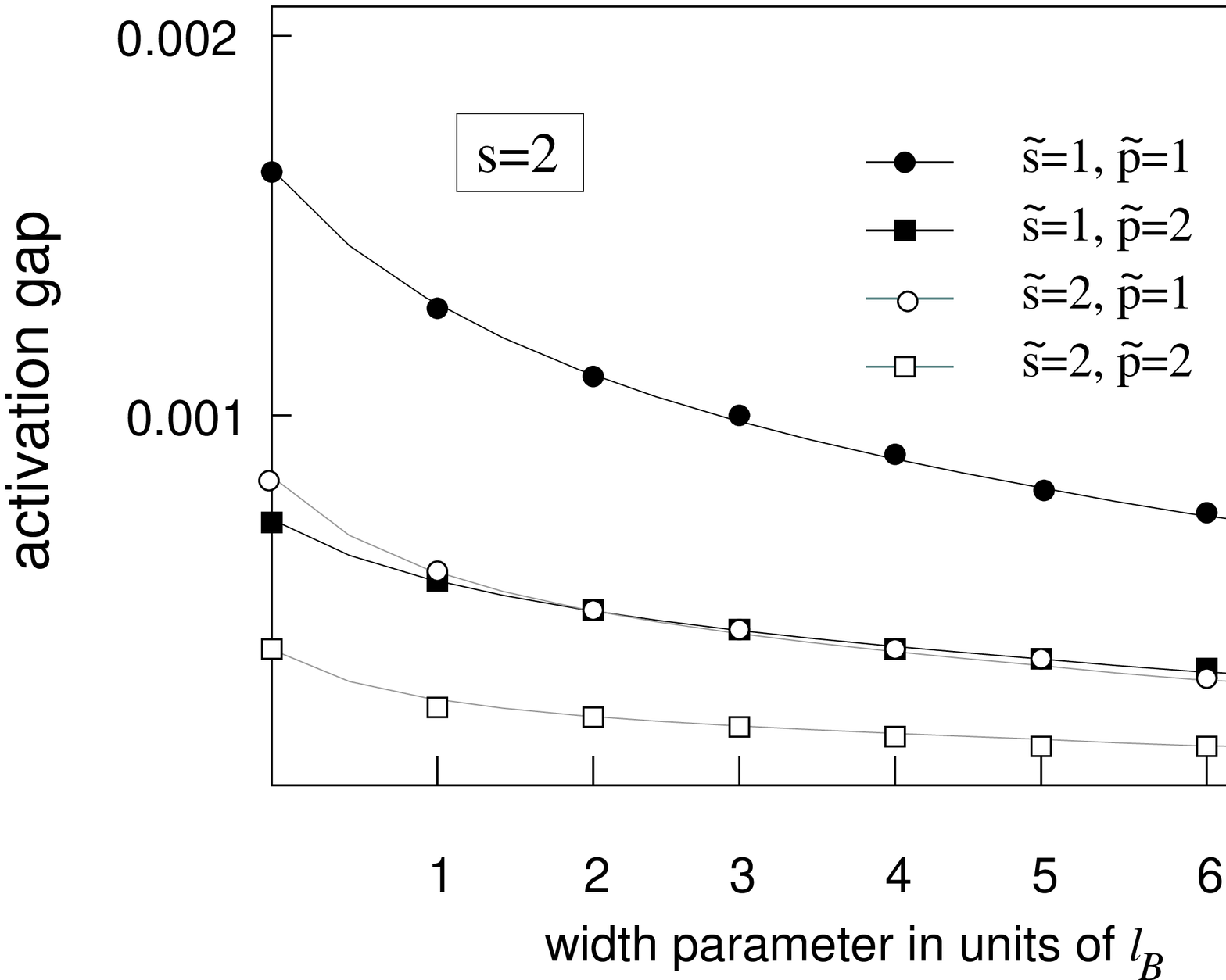}
\caption{Activation gaps as a function of width parameter for different fillings. (a) $s=1$: $\nu=4/11$ (black circles), $\nu=6/17$ (white circles), $\nu=7/19$ (black squares), and $\nu=11/31$ (white squares); (b) $s=2$:  $\nu=4/19$ (black circles), $\nu=6/29$ (white circles), $\nu=7/33$ (black squares), and $\nu=11/53$ (white squares).}
\label{fig01}
\end{figure}

Finite width effects may be included by replacing the original
interaction potential $v_0(q)\rightarrow v_0(q)f(\lambda,q)$, where
$f(\lambda,q)=\exp(\lambda^2q^2)[1-{\rm Erf}(\lambda q)]$ is given in terms
of the error function ${\rm Erf}(x)$. This expression has been obtained under 
the assumption that the confining potential in the $z$-direction (with a
characteristic width $\lambda$) be quadratic. \cite{MS,morf} The
results are shown in Fig.\,\ref{fig01} for different filling
factors. As expected from the FQHE of electrons, the activation gap is
reduced for samples with larger $\lambda$. \cite{morf}

The only activation gap of this series which was inferred from 
experiments is the
one at $\nu=4/11$, which is on the order of $50$mK for a sample width
of $\sim 500$\AA. \cite{pan} This corresponds to a width parameter
$\lambda\simeq 6.3 l_B$ at $B=10$T. The activation gap obtained
theoretically for this width is
$\Delta^a(s=1;\stilde=1,\ptilde=1)\simeq 0.0025 e^2/\epsilon l_B$,
which is on the order of $400$mK and thus one order of magnitude
larger than the experimental activation gap and almost three times
larger than the value calculated by Chang {\sl et al.} for a partially
spin-polarized state.\cite{chang} Two effects may account 
for this discrepancy. First, it is
known that the Hamiltonian theory for the conventional FQHE
overestimates the activation gap by a factor of $\sim1.5$ in
comparison with numerical studies even if the agreement becomes better
for wider samples. \cite{MS} This effect is expected
to be also present in the C$^2$F formalism. A second and more likely
origin of the discrepancy between the theoretical and experimental
results is the fact that the
theory does not account for impurity effects. They lead to a gap
reduction, which may be on the same order
of magnitude as the activation gap in the absence of impurities.
\cite{morf2} Impurities affect these sensitive C$^2$F states more
than their corresponding CF states because the activation gap of the
latter is more than one order of magnitude larger. 

\begin{figure}
\epsfysize+5.0cm
\epsffile{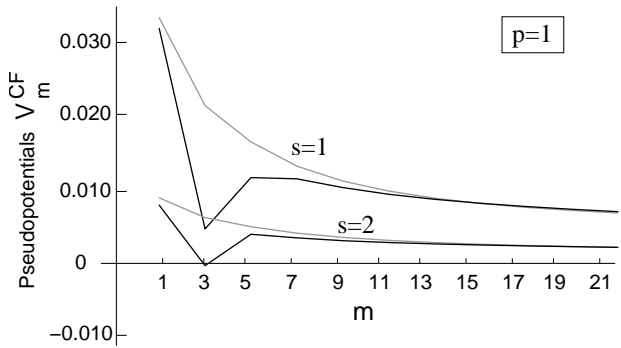}
\caption{Haldane's pseudopotentials for CF interaction both without screening
(black line) and with screening (gray line); the lines are a guide to the 
eyes.}
\label{fig02}
\end{figure}

\section{Criterion for the stability of C$^2$Fs}

Finally, we comment on the stability of C$^2$Fs. In order to conclude about 
their existence in nature, one would in principle have
to compare the energy of the C$^2$F liquid with possibly competing phases,
{\sl e.g.} solid phases of CFs in $p=1$. \cite{leejain} 
However, insight into their stability may 
be obtained from an expansion of the CF interaction potential (\ref{equ012}) 
in Haldane's pseudopotentials, \cite{haldane}
$V_{m}^{CF}=\sum_{\bq}\vtilde(q)L_{m}(q^2l_B^{2*})\exp(-q^2l_B^{2*}/2)$,
with odd $m$. The results for $p=1$ and different $s$, {\sl i.e.} for the 
CF interaction in the filling-factor range $1/(2s+1)<\nu<2/(4s+1)$, are shown 
in Fig.\,\ref{fig02}. In contrast to the electronic case, one observes a 
non-monotonic behavior for the unscreened CF interaction (black lines); the 
largest value is found for $m=1$, but there 
is a large suppression of the component $V_3^{CF}$, which is smaller than 
$V_5^{CF}$ and even becomes {\sl negative} for larger $s$. A similar result
has been obtained in previous studies in the 
wave-function approach. \cite{beran,leejain} The non-monotonic behavior is 
likely to be due to the dipole-like internal structure of the CF. 
\cite{MS,read,PH} 
An attractive dipole-dipole interaction is thus superimposed on the repulsive 
Coulomb repulsion. This stabilizes the C$^2$F states at $\nu^*=1+1/3$, such as
the $4/11$ state, which have their largest weight in the $m=1$ 
angular-momentum component, but may destabilize states at $\nu^*=1+1/5$, 
{\sl e.g.} the one at $\nu=6/17$. If one takes into account the screening, 
given by the dielectric function (\ref{equ020}), the component $V_3^{CF}$ 
increases, and the potential becomes more Coulomb-like (gray lines). 
The $4/11$ state is found to remain stable in the presence of screening
because for $s=1$ the ratio $V_1^{CF}/V_3^{CF}=1.56$ remains 
sufficiently high, {\sl i.e.} the potential is sufficiently short-range.
A first-order phase transition to a different state is 
expected to appear for a ratio $V_1/V_3\lesssim 1.2$. \cite{quantumhall}

\section{Conclusions}

In conclusion, we have interpreted the experimentally observed FQHE \cite{pan} 
at $\nu=4/11$ as an IQHE of a second generation of CFs (C$^2$Fs)
in the framework of a recently developed model of interacting CFs, 
\cite{goerbigUP} which 
is based on the Hamiltonian theory of the FQHE. \cite{MS}
After restriction of the dynamics to the CF-LL $p=1$, the
effective Hamiltonian for C$^2$Fs becomes similar
to the CF Hamiltonian in the lowest electronic LL. The
formalism is generic and may be applied for other CF-LLs, $p>1$, as well as for
higher generations of CFs. \cite{goerbigUP} Screening effects due to 
inter-CF-LL excitations,
which in contrast to their electronic counterpart cannot be
neglected even in the large-$B$ limit, are included in a dielectric
function, calculated in the RPA. Both
screening and the finite width of the two-dimensional electron gas
give rise to a modification of the effective interaction potential and
lead to a reduction of the C$^2$F activation gaps, which are about one
order of magnitude smaller than those for CFs. The theoretical gap at 
$\nu=4/11$ is, however, substantially larger than the gap inferred from the
experiment. \cite{pan} This discrepancy is likely to be due to 
residual impurities in the sample, which have been omitted in the
theoretical model and which are known to produce a further reduction of the
activation gaps. \cite{morf2} Whereas the
stability of a fully polarized 4/11 state remains controversial,
\cite{beran,wojs,mandaljain,lopezfradkin,haldane2} a pseudopotential
expansion of the CF interaction potential hints to a stable 4/11 state
if interpreted in terms of C$^2$Fs. The definitive answer to this
question, however, requires detailed energy calculations involving all
possible competing states, namely CF Wigner crystal, CF bubble, and
C$^2$F liquid phases.

We acknowledge fruitful discussions with R.\ Morf. This work was supported 
by the Swiss National Foundation for
Scientific Research under Grant No.~620-62868.00.

\newpage

\centerline{\bf Erratum}

A factor $n_B^*=1/2\pi l_B^{*2}$ is missing in Eq. (8). Its corrected version 
is
$$
\mathcal{F}_n^p(q)=n_B^*\sum_{n'=p-n}^{p-1}|\langle
n'+n| \rhobar^p(\bq)| n'\rangle|^2.
$$
This error has led to an overestimation of the screening effect due to 
inter-CF-LL
excitations. Consequently screening affects the values of the activation gaps
less than originally thought. The corrected values for the gaps with $s=1$ 
and $p=1$

\begin{center}
\begin{tabular}{|c||c|c|}
\hline
$\stilde=1$ & $\ptilde=1$ & $\ptilde=2$  \\ \hline
$\nu^*$ & $1+1/3$ & $1+2/5$ \\ \hline
$\nu$ & $4/11$ & $7/19$  \\ \hline \hline
$\Delta^a$ & 0.018 & 0.0064  \\ \hline
\end{tabular}\hspace{0.5cm}
\begin{tabular}{|c||c|c|}
\hline
$\stilde=2$ & $\ptilde=1$ & $\ptilde=2$  \\ \hline
$\nu^*$ & $1+1/5$ & $1+2/9$  \\ \hline
$\nu$ & $6/17$ & $11/31$ \\ \hline \hline
$\Delta^a$ & 0.011 & 0.0052 \\ \hline
\end{tabular}
\end{center}

and for $s=2$ and $p=1$

\begin{center}
\begin{tabular}{|c||c|c|}
\hline
$\stilde=1$ & $\ptilde=1$ & $\ptilde=2$ \\ \hline
$\nu^*$ & $1+1/3$ & $1+2/5$ \\ \hline
$\nu$ & $4/19$ & $7/33$ \\ \hline \hline
$\Delta^a$ & 0.0057 & 0.0020 \\ \hline
\end{tabular}\hspace{0.5cm}
\begin{tabular}{|c||c|c|}
\hline
$\stilde=2$ & $\ptilde=1$ & $\ptilde=2$ \\ \hline
$\nu^*$ & $1+1/5$ & $1+2/9$ \\ \hline
$\nu$ & $6/29$ & $11/53$ \\ \hline \hline
$\Delta^a$ & 0.0041 & 0.0021  \\ \hline
\end{tabular}
\end{center}

\noindent
in units of $e^2/\epsilon l_B$.
\begin{figure}
\epsfysize+4.5cm
\epsffile{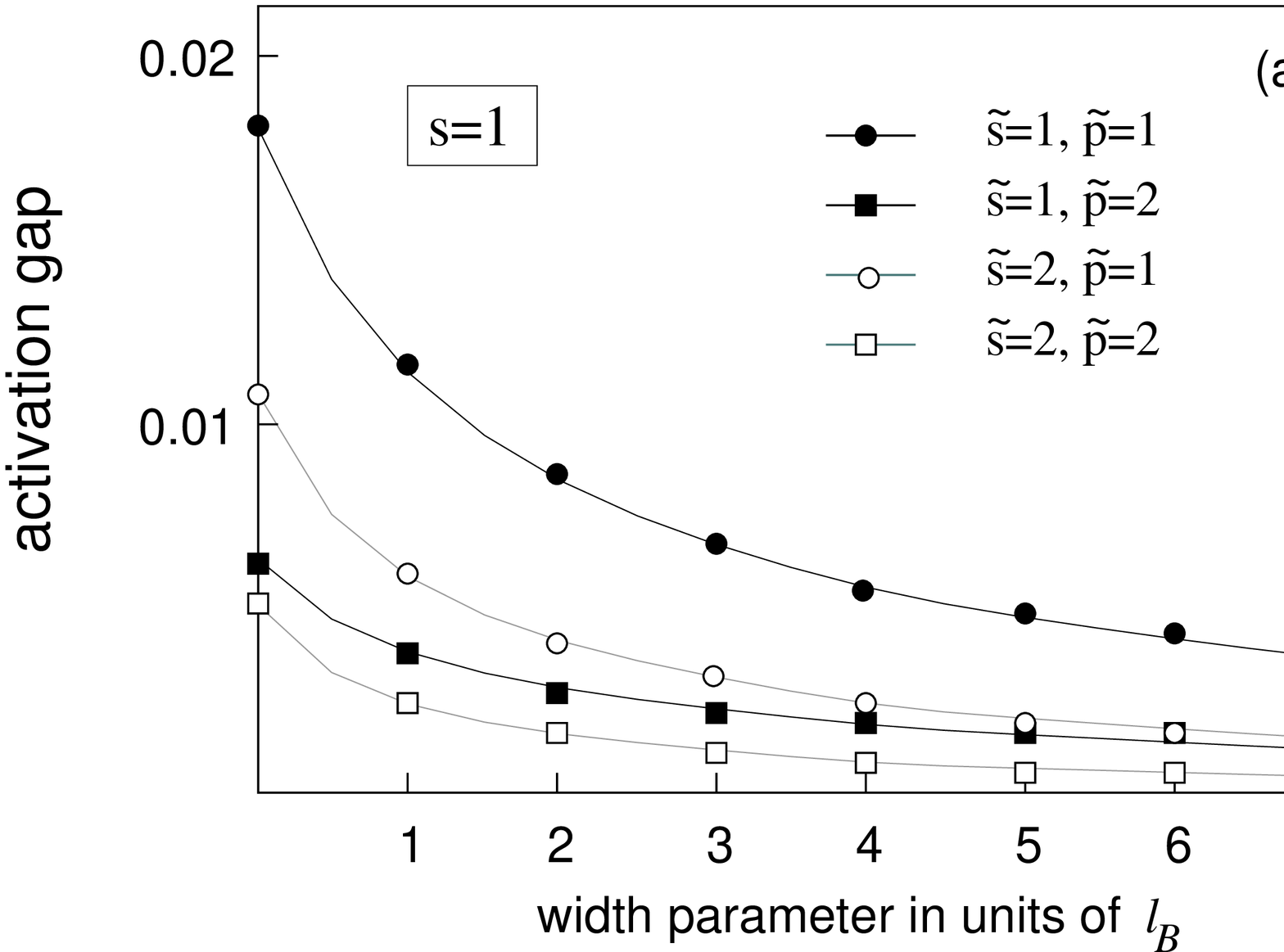}
\epsfysize+4.5cm
\epsffile{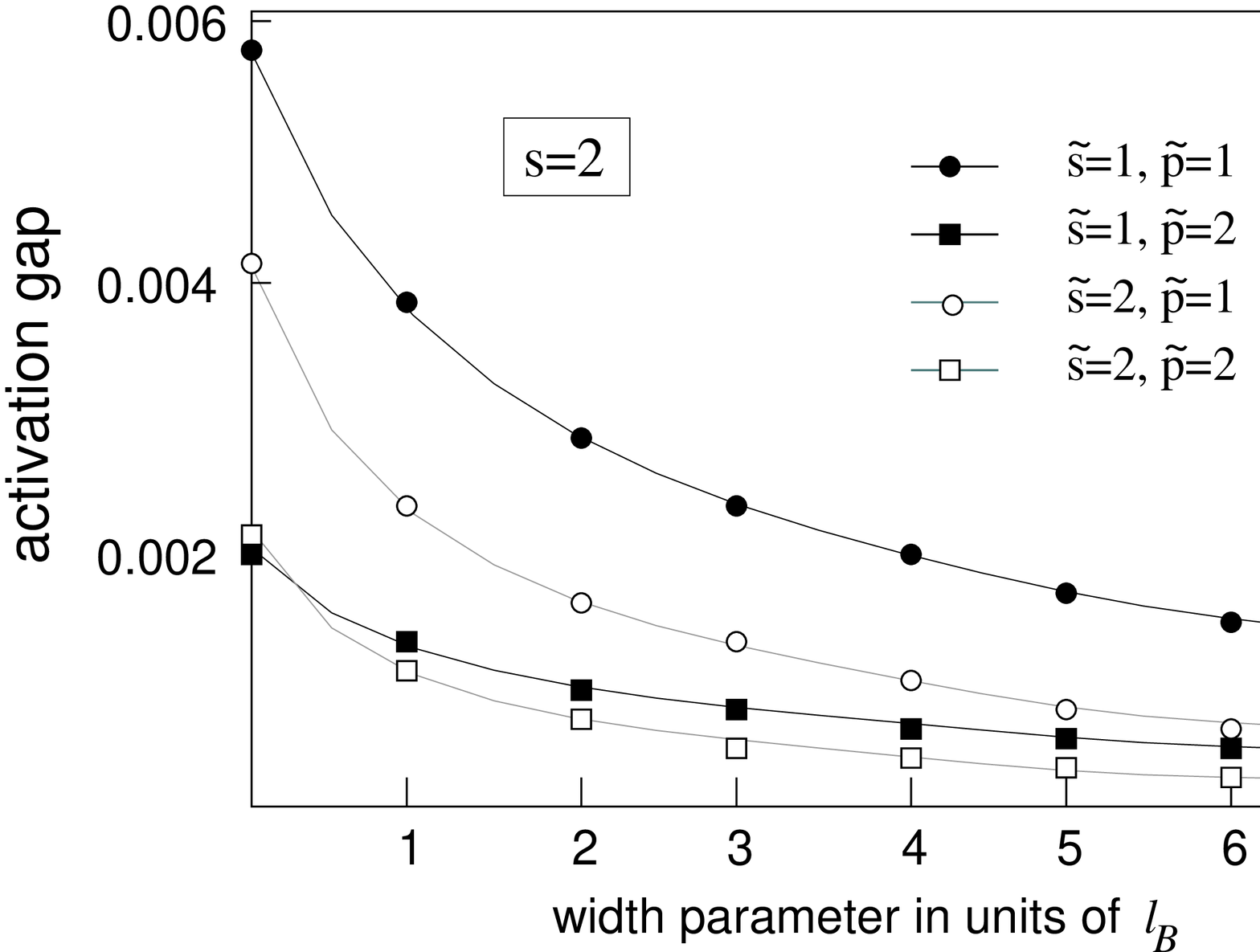}
\caption{Activation gaps.
(a) $s=1$: $\nu=4/11$ (black circles), $\nu=6/17$ (white circles), 
$\nu=7/19$ (black squares), and $\nu=11/31$ (white squares); (b) $s=2$:  
$\nu=4/19$ (black circles), $\nu=6/29$ (white circles), $\nu=7/33$ (black 
squares), and $\nu=11/53$ (white squares).}
\label{fig01err}
\end{figure}

\begin{figure}
\epsfysize+4.5cm
\epsffile{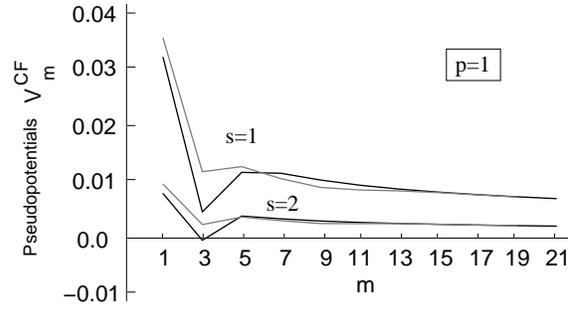}
\caption{Haldane's pseudopotentials for CF interaction both without screening
(black line) and with screening (gray line); the lines are a guide to the 
eyes.}
\label{fig02err}
\end{figure}

The corrected figures may be found in Figs. \ref{fig01err} and \ref{fig02err}.
The theoretical estimate for the activation gap at $\nu=4/11$ for a width 
parameter $\lambda\simeq 6.3l_B$ is 
$\Delta^a(s=1;\stilde=\ptilde=1)\simeq 0.004 e^2/\epsilon l_B$, which is on
the order of $600$mK.
The characteristic minimum of $V_3^{CF}$ persists even in the presence
of screening (Fig. \ref{fig02err}).


\begin{thebibliography}{99}

\bibitem{jain}J.\ K.\ Jain, Phys.\ Rev.\ Lett. {\bf 63}, 199 (1989);
  Phys.\ Rev.\ B\ {\bf 41}, 7653 (1990).

\bibitem{perspectives}S.\ Das\ Sarma and A.\ Pinczuk, eds., {\sl
    Perspectives in Quantum Hall Effects}, Wiley, New York (1997).

\bibitem{pan}W.\ Pan, H.\ L.\ Stormer, D.\ C.\ Tsui, L.\ N.\ Pfeiffer, 
K.\ W.\ Baldwin, and K.\ W.\ West, Phys.\ Rev.\ Lett.\ {\bf 90}, 016801 (2003).

\bibitem{mani}R.\ G.\ Mani and K.\ v.\ Klitzing, Z.\ Phys.\ B\ {\bf
    100}, 635 (1996). 

\bibitem{laughlin}R.\ B.\ Laughlin, Phys.\ Rev.\ Lett.\ {\bf 50}, 1395 (1983).

\bibitem{haldane}F.\ D.\ Haldane, Phys.\ Rev.\ Lett.\ {\bf 51}, 605 (1983).

\bibitem{halperin}B.\ I.\ Halperin, Phys.\ Rev.\ Lett.\ {\bf 52}, 1583 (1984).

\bibitem{beran}P.\ B\'eran and R.\ Morf, Phys.\ Rev.\ B\ {\bf 43}, 12654. 
(1991).

\bibitem{wojs}A.\ W\'ojs and J.\ J.\ Quinn, Phys.\ Rev.\ B\ {\bf 61}, 2846 
(2000).

\bibitem{mandaljain}S.\ S.\ Mandal and J.\ K.\ Jain, Phys.\ Rev.\ B\ {\bf 66}, 
155302 (2002).

\bibitem{chang}C.-C.\ Chang, S.\ S.\ Mandal, and J.\ K.\ Jain, Phys.\
  Rev.\ B\ {\bf 67}, 121305 (2003).

\bibitem{morfPC}R.\ Morf, {\sl private communication}.

\bibitem{lopezfradkin}A.\ Lopez and E.\ Fradkin, cond-mat/0310128.

\bibitem{haldane2}F.\ D.\ Haldane, Phys.\ Rev.\ Lett.\ {\bf 74}, 2090 (1995).

\bibitem{goerbigUP}M.\ O.\ Goerbig, P.\ Lederer, and C.\ Morais\
  Smith, cond-mat/0401340.

\bibitem{MS}G.\ Murthy and R.\ Shankar, Rev.\ Mod.\ Phys.\ {\bf 75}, 1101 
(2003); R.\ Shankar, Phys.\ Rev.\ B\ {\bf 63}, 085322 (2001). 

\bibitem{AG}I.\ L.\ Aleiner and L.\ I.\ Glazman, Phys.\ Rev.\ B\ {\bf
    52}, 11296 (1995). 

\bibitem{GMP}S.\ M.\ Girvin, A.\ H.\ MacDonald, and P.\ M.\ Platzman,
  Phys.\ Rev.\ B\ {\bf 33}, 2481 (1986). 

\bibitem{morf}R.\ Morf, N.\ d'Ambrumenil, and S.\ Das Sarma, Phys.\
  Rev.\ B\ {\bf 66}, 075408 (2002).

\bibitem{morf2}R.\ Morf and N.\ d'Ambrumenil, Phys.\ Rev.\ B\ {\bf 68}, 
113309 (2003).

\bibitem{leejain}S.-Y.\ Lee, V.\ W.\ Scarola, and J.\ K.\ Jain, Phys.\ Rev.\ 
Lett.\ {\bf 87}, 256803 (2001); Phys.\ Rev.\ B\ {\bf 66}, 085336 (2002).

\bibitem{read}N.\ Read, Semi.\ Sci.\ Tech.\ {\bf 9}, 1859 (1994).

\bibitem{PH}V.\ Pasquier and F.\ D.\ Haldane, Nucl.\ Phys.\ B {\bf
    516}, 719 (1998).

\bibitem{quantumhall}D.\ Yoshioka, {\sl The Quantum Hall Effect}, Springer, 
Berlin (2002).

\end{thebibliography}
\end{document}